\documentclass{article}

\usepackage{PRIMEarxiv}
\usepackage{amsmath}
\usepackage{subcaption}

\usepackage[utf8]{inputenc} 
\usepackage[T1]{fontenc}    
\usepackage{hyperref}       
\usepackage{url}            
\usepackage{booktabs}       
\usepackage{amsfonts}       
\usepackage{nicefrac}       
\usepackage{microtype}      
\usepackage{lipsum}
\usepackage{graphicx}
\graphicspath{{media/}}     

\title{Automated Repair of C Programs Using Large Language Models
}

\author{
  Mahdi Farzandway, Fatemeh Ghassemi \\
  University of Tehran \\
  \texttt{\{mahdifarzandway, fghassemi\}@ut.ac.ir} \\
}

\begin{document}
\maketitle

\begin{abstract}
This study explores the potential of Large Language Models (LLMs) in automating the repair of C programs. We present a framework that integrates spectrum-based fault localization (SBFL), runtime feedback, and Chain-of-Thought–structured prompting into an autonomous repair loop. Unlike prior approaches, our method explicitly combines statistical program analysis with LLM reasoning. The iterative repair cycle leverages a structured Chain-of-Thought (CoT) prompting approach, where the model reasons over failing tests, suspicious code regions, and prior patch outcomes, before generating new candidate patches. The model iteratively changes the code, evaluates the results, and incorporates reasoning from previous attempts into subsequent modifications, reducing repeated errors and clarifying why some bugs remain unresolved. Our evaluation spans $3,902$ bugs from the Codeflaws benchmark, where our approach achieves $44.93\%$ repair accuracy—representing a $3.61\%$ absolute improvement over strong state-of-the-art APR baselines such as GPT-4 with CoT. This outcome highlights a practical pathway toward integrating statistical program analysis with generative AI in automated debugging.
\end{abstract}


\section{Introduction}

In software development, debugging remains one of the most challenging and time-intensive tasks, particularly as codebases grow in size and complexity. The process of identifying and correcting programming bugs requires not only expertise but also significant manual effort. Traditional debugging methods typically rely on static or dynamic analysis tools to assist in bug detection. However, these methods often fail to fully automate the correction process, requiring developers to analyze the root cause of errors, apply fixes, and validate solutions through testing.

The advent of Artificial Intelligence (AI), particularly the rise of \textit{Large Language Models (LLMs)}, has brought a transformative shift to this domain. LLMs, such as GPT-3 \cite{brown2020language}, Codex \cite{chen2021evaluating}, and CodeT5 \cite{wang2021codet5}, trained on extensive datasets of natural language and code, are capable of generating and reasoning about code. This capability makes them valuable tools for automating tasks like code completion, bug detection, and program repair. However, debugging presents unique challenges compared to tasks such as code generation due to its iterative nature, where multiple cycles of identification, correction, and validation are required.

Studies indicate that developers devote approximately $48\%$ of their programming time to identifying and fixing bugs \cite{alaboudi2021exploratory}. The emergence of LLMs has significantly transformed various aspects of software engineering, particularly in the realm of automated bug detection and correction \cite{chen2021evaluating, xu2024automated}. Unlike traditional debugging methods that heavily depend on manual intervention and rule-based systems, LLMs introduce a paradigm shift in software maintenance and reliability \cite{zhang2024autocoderover, xia2023keep}.

Recent advances in LLM-based program repair have demonstrated their ability to comprehend complex code structures and generate contextually relevant fixes~\cite{ruiz2024novel, le2024study}. However, challenges persist in ensuring consistency and generalizability across diverse programming contexts, such as different languages, frameworks, and project requirements~\cite{lyu2024automatic}. Integrating LLMs with established software engineering practices like test-driven development (TDD) and spectrum-based fault localization (SBFL) offers a promising avenue to address these challenges by leveraging test case information and program context more effectively for automated program repair (APR)~\cite{kong2024contrastrepair, xia2023conversational}. Despite these advancements, fundamental issues remain in automated debugging systems, particularly with complex software defects that demand extensive domain expertise or reasoning beyond current LLM capabilities~\cite{allamanis2018survey}. While LLMs excel in recognizing patterns and common errors, they struggle with novel or domain-specific problems. 
The effectiveness of LLMs in debugging is also heavily reliant on the quality of feedback from sources like test results or fault localization techniques. Integrating LLMs with iterative feedback mechanisms and advanced fault localization strategies, as well as hybrid approaches combining LLMs with static analysis and symbolic execution, shows significant potential for overcoming these limitations and enhancing the precision and efficiency of APR tools.

Our research addresses several critical limitations in current automated debugging approaches. First, existing solutions often struggle with the accuracy-efficiency trade-off, either providing quick but unreliable fixes or accurate but computationally expensive solutions \cite{hajipour2021samplefix, wei2023copiloting}. Second, many approaches cannot leverage multiple sources of debugging information effectively, potentially missing crucial insights that could lead to more robust solutions \cite{drain2021generating, jiang2023knod}. Third, the challenge of maintaining semantic consistency while generating fixes remains largely unaddressed in the current literature \cite{duvsek2020evaluating}.

To address these challenges, we present a framework that integrates an autonomous \emph{AI Debugger Agent} with an iterative refinement process, leveraging test-case results and spectrum-based fault-localization (SBFL) analysis. Our design employs an external agent loop that repeatedly provides the LLM with structured feedback composed of failing and passing test cases, SBFL-based rankings of suspicious code locations, and relevant runtime warnings. In each iteration, the agent and LLM jointly execute a cycle of \emph{[analyze error] $\rightarrow$ [analyze previous tries] $\rightarrow$ [hypothesize fix] $\rightarrow$ [generate patch] $\rightarrow$ [re-test]}, where the arrows denote sequential handoffs between iterations rather than reasoning steps within a single model response. This operationalization enables stepwise reasoning across multiple interactions. We emphasize that our approach implements \emph{iterative reasoning}, rather than in-context Chain-of-Thought (CoT) prompting, because the latter restricts reasoning to a single pass of model-generated tokens and cannot effectively incorporate dynamic runtime feedback or insights from prior unsuccessful attempts.

Our approach integrates test-case results, SBFL scores, and runtime warnings into a structured iterative refinement framework for APR. While prior works have explored individual feedback mechanisms, our contribution is the systematic unification of multiple feedback sources within a single repair loop. Specifically, we combine test-case outcomes, suspiciousness scores, and runtime warnings into structured prompts that guide the LLM through stepwise reasoning across iterations. Unlike existing methods, which often rely on a single feedback type or employ ad-hoc iterative strategies, our framework maintains an explicit memory of previous attempts; this memory enables the LLM to avoid repeating ineffective changes and progressively refine patches. This design explicitly addresses three key limitations of prior approaches: (i) dependence on a single feedback signal, (ii) absence of principled iterative refinement, and (iii) lack of a mechanism to leverage information from previous repair attempts to guide future iterations.

This methodology represents a significant advancement over existing approaches, achieved through the following key features:
\begin{itemize}
    \item \textbf{Incorporating Multiple Feedback Mechanisms}: Our framework guides the debugging process by integrating test-case results, SBFL, and runtime warnings---messages generated during code execution that indicate potential issues. These warnings, along with test-case outcomes and SBFL, are provided to the language model (LLM) at each stage, enabling the LLM to refine the code for improved bug detection and correction iteratively.
    \item \textbf{Iterative Refinement Strategy}: We implement an iterative refinement strategy that progressively enhances solution quality while maintaining computational efficiency, enabling the continuous improvement of code until optimal results are achieved.
    \item \textbf{Utilizing Advanced Prompt Engineering Techniques}: Our approach enhances the reasoning capabilities and solution accuracy of LLMs by incorporating information from previous repair attempts, enabling the system to avoid repeating past errors and produce contextually appropriate fixes.
    \item \textbf{Comprehensive Evaluation Framework}: A thorough evaluation framework is developed to assess both the quality and reliability of generated fixes, providing a robust basis for validating the effectiveness of our approach.
\end{itemize}

These features collectively enhance the precision, efficiency, and reliability of our LLM-based automated program repair tool, setting a new standard for automated debugging solutions. Our evaluation spans $3,902$ bugs from the Codeflaws benchmark, where our approach achieves $44.93\%$ repair accuracy—representing a $3.61\%$ absolute improvement over strong state-of-the-art APR baselines such as GPT-4 with CoT. These results not only establish the effectiveness of our methodology but also provide insights into the potential of combining statistical program analysis with generative AI for enhanced automated debugging workflows.

The remainder of this paper is organized as follows: Section \ref{Sec::Back} 
provides comprehensive background information. Section \ref{Sec::Method} 
details our methodology and implementation. Section \ref{Sec::Exp} 
presents our experimental results, followed by a summary of the results in Section \ref{Sec::Res}. 
Section \ref{Sec::Related}
discusses related work in this domain. Section \ref{Sec::Lim} 
explores the limitations and future research directions, and finally, Section \ref{Sec::Con} 
concludes the paper.

\section{Background\label{Sec::Back}}
In this study, we discuss two key components that underpin our approach to automated program repair: (1) \textit{Chain of Thought} prompting, which enables LLMs to reason step-by-step when generating candidate patches, and (2) \textit{Spectrum-Based Fault Localization (SBFL)}, which helps focus repair efforts on suspicious code regions. While both techniques have been explored in prior work, our contribution lies in their integration into a unified repair pipeline. To situate our work, we also highlight limitations of existing baselines and explain how our method specifically addresses them.

\subsection{Chain of Thought}
\label{subsec::ChainOfThought}
Large Language Models (LLMs) have recently emerged as a transformative technology in automated debugging, demonstrating strong capabilities in code interpretation and error correction \cite{chen2021evaluating}. They can analyze and repair code snippets, often surpassing traditional static analysis tools in detecting subtle logical errors, and their ability to generate context-aware corrections has opened new possibilities for automated programming assistance.  

Within this broader paradigm, one of the most influential prompting strategies is \textit{Chain-of-Thought (CoT)} prompting. CoT is a prompt-engineering technique that elicits intermediate reasoning steps from a model instead of asking it to directly produce a final answer. By encouraging the model to generate a sequence of explicit inference steps, CoT improves performance on multi-step reasoning tasks such as arithmetic, logical reasoning, and multi-hop inference. Common variants include \textit{few-shot CoT}, where exemplars with worked-out solutions are provided in the prompt, and \textit{zero-shot CoT}, where a short instruction (e.g., ``Let’s think step by step’’) is sufficient to induce reasoning traces. More advanced methods, such as \textit{self-consistency}, combine multiple reasoning chains to increase robustness. While CoT has been shown to significantly boost correctness on complex reasoning benchmarks, it also introduces trade-offs: longer outputs (hence higher computational cost), sensitivity to prompt phrasing, and the risk of producing plausible but incorrect reasoning.  

In our work, we leverage CoT’s ability to decompose reasoning into intermediate steps for program repair. Specifically, our prompts instruct the model to (i) explain the observed failure step by step, (ii) identify suspicious statements or expressions, and (iii) propose a minimal candidate patch with justification. This structured “reason-then-repair’’ flow reduces blind guessing and yields more interpretable candidate patches compared to direct patch-generation prompts.

A key advancement in this domain is the use of iterative feedback processes. In our approach, \textit{Chain of Thought} (CoT) prompting is explicitly operationalized in the repair workflow by structuring stepwise prompts as "\emph{[analyze error] $\rightarrow$ [analyze previous tries] $\rightarrow$ [hypothesize fix] $\rightarrow$ [generate patch] $\rightarrow$ [re-test]}." This design distinguishes our method from prior iterative repair loops, where the reasoning process remained external to the model. By embedding reasoning steps directly into the LLM prompt, our approach ensures both interpretability and stronger alignment between failure analysis and patch generation.

Through iterative examination of test case outcomes, the LLM incrementally refines its corrections until the code passes all tests or meets the required functionality. This process mirrors human debugging workflows, where developers repeatedly analyze failures, analyze previous tries, hypothesize fixes, apply adjustments, and re-test until all errors are resolved.

\subsection{Spectrum-Based Fault Localization (SBFL)} 
\label{subsec::SBFL}
To complement LLM-driven program repair, we integrate Fault localization (FL) results through \textit{Spectrum-Based Fault Localization (SBFL)}. 
Fault localization (FL) is a technique for identifying likely buggy code locations. 
However, prior APR frameworks often use FL only as auxiliary input or ignore it altogether. 
In contrast, our method tightly couples SBFL suspiciousness scores with the LLM’s reasoning process. SBFL is a dynamic analysis method that calculates the likelihood of specific code segments being faulty based on their execution behavior across multiple test cases. Widely adopted statistical approaches \cite{abreu2007accuracy, abreu2006evaluation, jones2005empirical} utilize test execution data to compute a \textit{suspiciousness score} for each statement, indicating its likelihood of being the origin of the fault.

\begin{figure*}
\centerline{\includegraphics[scale=0.33]{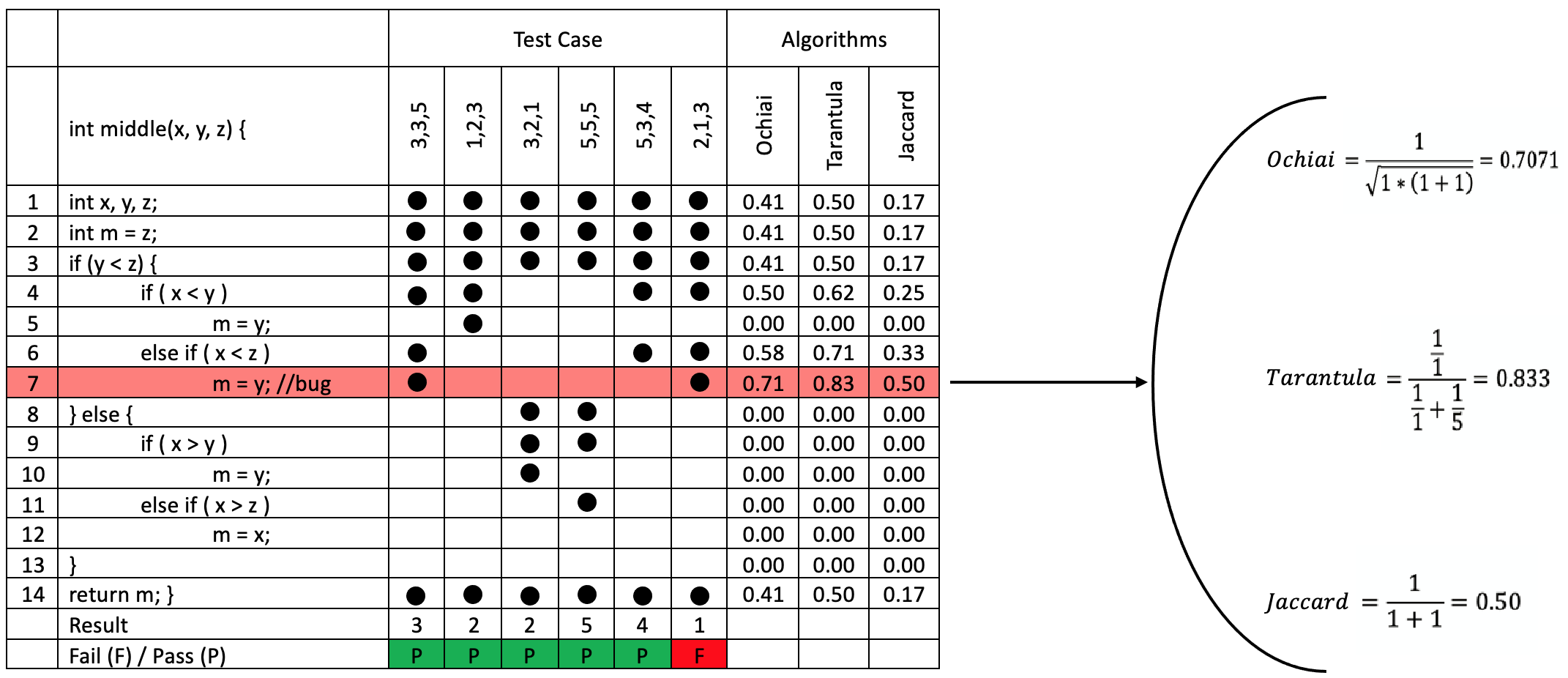}}
\caption{Application of different suspiciousness ranking formulas: black circles indicate executed statements within each test case \cite{farzandway2024specnlp}.}
\label{fig::spectrum}
\end{figure*}

These approaches rely on metrics such as the total number of failed and passed test cases, as well as the number of tests that executed specific statements. Common formulas, such as Jaccard \cite{abreu2007accuracy}, Ochiai \cite{abreu2006evaluation}, and Tarantula \cite{jones2005empirical}, are employed to rank the suspiciousness of statements:

\[
\begin{array}{c}
\text{Jaccard: } \frac{T_f(e)}{T_f + T_p(e)} \quad \text{Ochiai: } \frac{T_f(e)}{\sqrt{T_f \times (T_f(e) + T_p(e))}} \\
\text{Tarantula: } \frac{\frac{T_f(e)}{T_f}}{\frac{T_f(e)}{T_f} + \frac{T_p(e)}{T_p}}
\end{array}
\]where $T_f$ and $T_p$ represent the total number of failed and passed test cases, respectively, while $T_f(e)$ and $T_p(e)$ denote the number of failed and passed test cases in which a statement $e$ is executed. The statement with the highest suspiciousness score is considered the most likely location of the fault. Fig \ref{fig::spectrum} illustrates the application of these formulas.

By integrating SBFL with LLMs, our approach directly addresses a limitation of prior LLM-based repair methods, which often wasted iterations exploring irrelevant code. The suspiciousness scores are injected into the CoT-structured prompts, guiding the LLM to reason explicitly about the most error-prone statements. This synergy reduces unnecessary search and improves repair efficiency.

\section{Methodology\label{Sec::Method}}
This study presents a methodology that integrates fault localization and automated repair within the software debugging process using Large Language Models (LLMs). 
We evaluate effectiveness using concrete metrics: repair accuracy, the number of iterative refinement steps (LLM calls), and time-to-repair. These metrics are assessed under different feedback types, including test-case results, Spectrum-Based Fault Localization (SBFL) scores, and runtime warnings. 
We combine SBFL with structured Chain-of-Thought (CoT) prompting to enhance the reasoning process behind code patch generation compared to prior methods. By leveraging test-case results, SBFL suspiciousness scores, and runtime warnings, the feedback loop guides the LLM toward generating more reliable bug fixes. 
Unlike previous feedback-driven frameworks, our approach explicitly incorporates fault-localization data into prompt refinement and applies CoT reasoning steps, leading to higher repair accuracy and reduced debugging iterations. 
The following subsections detail the key modules of this methodology, beginning with an overview of the AI Debugger Agent and its iterative refinement workflow.

\subsection{Iterative Program Repair Workflow}
The repair process follows an iterative refinement loop that integrates (i) Spectrum-Based Fault Localization (SBFL) suspiciousness scores, (ii) structured Chain-of-Thought (CoT) reasoning prompts (e.g., `analyze → hypothesize fix → generate corrected code`), and (iii) memory of prior failed attempts.
In each iteration, we collect concrete feedback items — failing test identifiers with expected vs. actual outputs, and per-statement SBFL suspiciousness scores (i.e., Ochiai score $> 0.5$ and corresponding code contexts). These signals are merged into a structured prompt (failing-tests → SBFL suspiciousness scores → runtime warnings → history) which guides the LLM's next patch proposal. The loop terminates when all tests pass or a predefined iteration limit is reached.

\begin{figure*}[ht]
    \centering
    \includegraphics[width=0.9\textwidth]{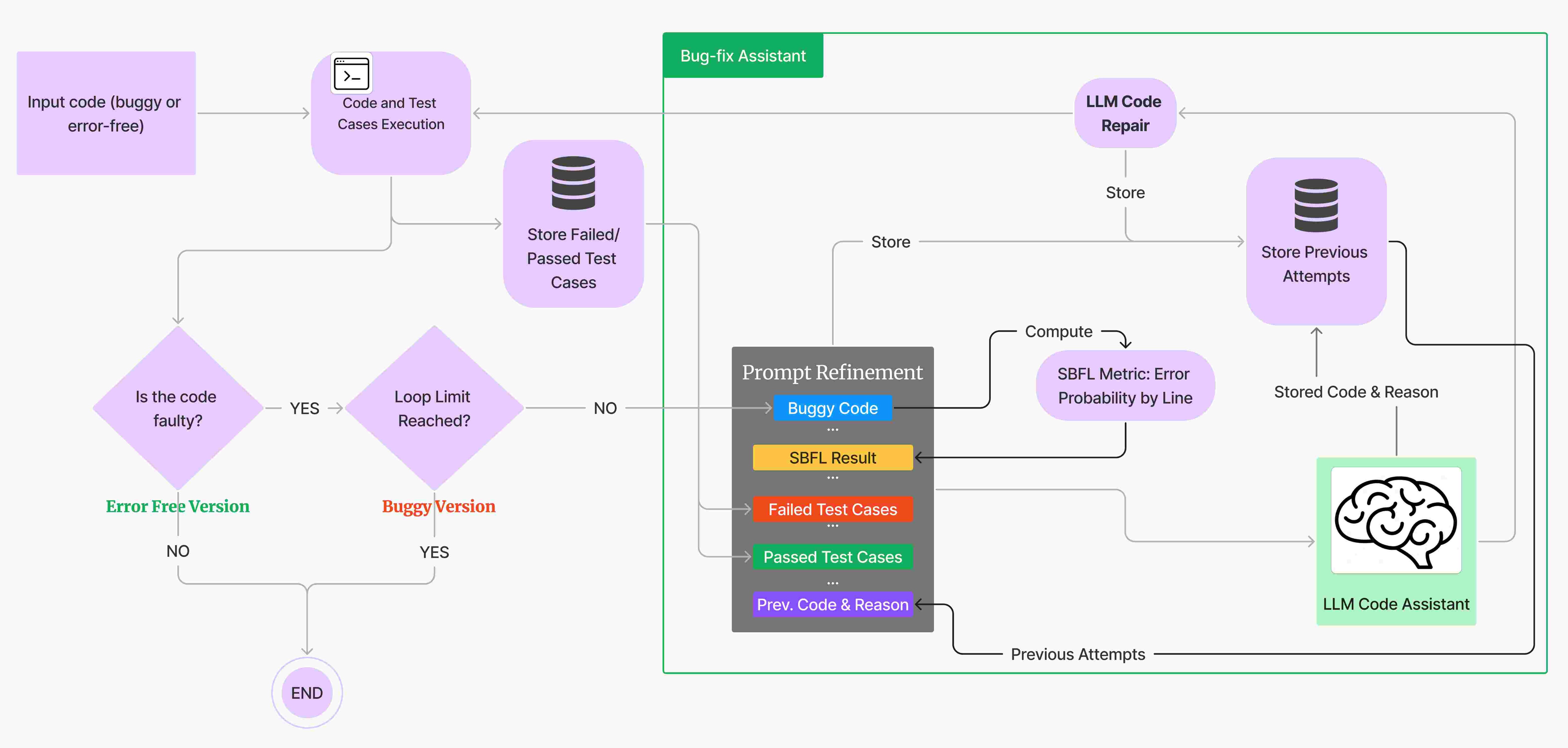}
    \caption{Iterative Refinement Workflow with LLM Integration}
    \label{fig:workflow}
\end{figure*}

The workflow as illustrated in Fig~\ref{fig:workflow}, consists of the following steps:

\begin{enumerate}
    \item \textbf{Code and Test Cases Execution:} The current code is executed against predefined test cases, logging pass/fail outcomes and error messages (the 'Code and Test Cases Execution' and 'Store Failed/Passed Test Cases' in Fig~\ref{fig:workflow}).

    \item \textbf{Feedback Collection:}
    Suspiciousness scores are calculated via SBFL based on the results of the test case executions. Following prior best practices~\cite{de2016spectrum, soremekun2021locating}, we compute suspiciousness with the Ochiai formula and filter and rank lines by their Ochiai scores, considering only the most probable faulty lines (i.e., Ochiai score $> 0.5$) as candidates for repair.
    These ranked lines, together with detailed test case outcomes, are then integrated into the LLM prompt, serving as structured input for the prompt refinement phase and enabling explicit CoT reasoning.

    \item \textbf{Prompt Refinement:} The LLM prompt is updated with SBFL results (highlighting the most suspicious lines), detailed outputs from failed test cases (including stack traces, expected vs observed outputs), and summaries of prior unsuccessful tries (hypotheses about fixes and previously generated patches). By integrating these elements with the reasoning steps, the process maintains a compact episodic memory retrieved in subsequent prompts, which helps the LLM refine its reasoning and avoid repeating previous mistakes.
    
    \item \textbf{Reiteration:} In the `reiteration' phase, the AI Debugger Agent uses the refined prompt to request an improved code version from the LLM. Our method then repeats the entire cycle (code and test cases execution, feedback collection, prompt refinement, and new code generation). This iterative process, driven by the AI Debugger Agent, continues until one of two conditions is met: either the execution results indicate that the code passes all test cases without any errors, or a predefined iteration limit (\textit{Loop Limit Reached}) is reached. Once either condition is satisfied, the debugging loop terminates.
\end{enumerate}

\begin{figure*}[ht]
    \centering
    \includegraphics[width=0.9\textwidth]{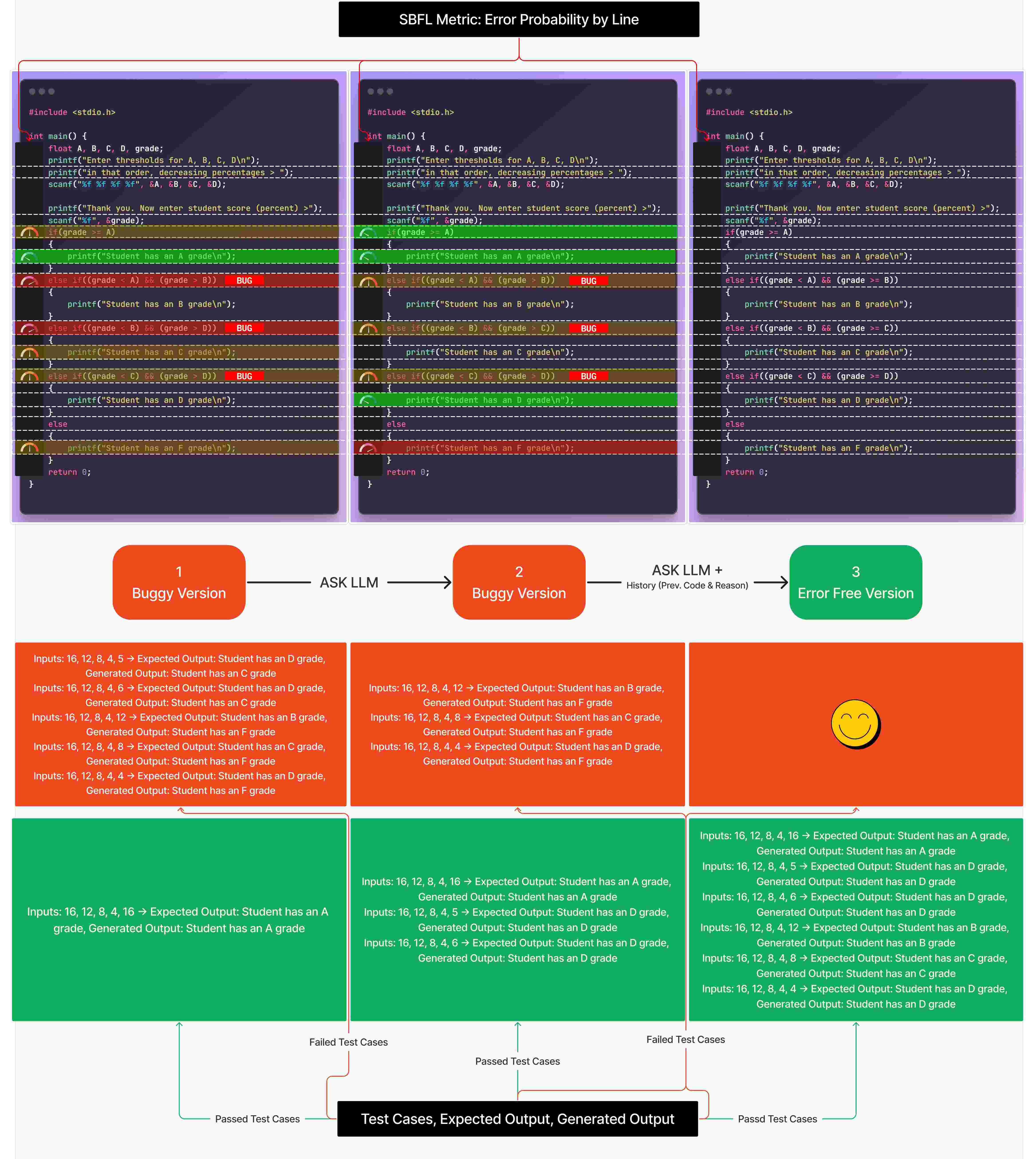}
    \caption{Code Debugging with SBFL, Suspiciousness Scoring, Test Case Feedback, and Generated Code History \& Reasoning}
    \label{fig:debugging_steps}
\end{figure*}

As illustrated in Fig \ref{fig:debugging_steps}, this structured approach involves iterative code execution where each cycle aims to resolve an increasing number of failed test cases. This progressive resolution ensures efficient and targeted debugging, minimizing redundant efforts until no test cases fail and error-free code is generated.

\subsection{Prompt Engineering for LLM Interaction}
Effective LLM interaction is vital for automated debugging. Recent studies highlight that well-designed prompt structures significantly enhance LLM-based code correction \cite{wong2023natural, shanuka2024systematic}, though prior work has not fully integrated spectrum-based fault localization (SBFL) and test feedback in a structured manner. Our methodology introduces a two-part prompt architecture that explicitly supports Chain-of-Thought reasoning, addressing prior limitations in clarity and systematic error analysis:

\begin{figure*}[t!]
	\centering
	\begin{subfigure}[t]{0.55\textwidth}
		\centering
		\includegraphics[width=0.8\textwidth]{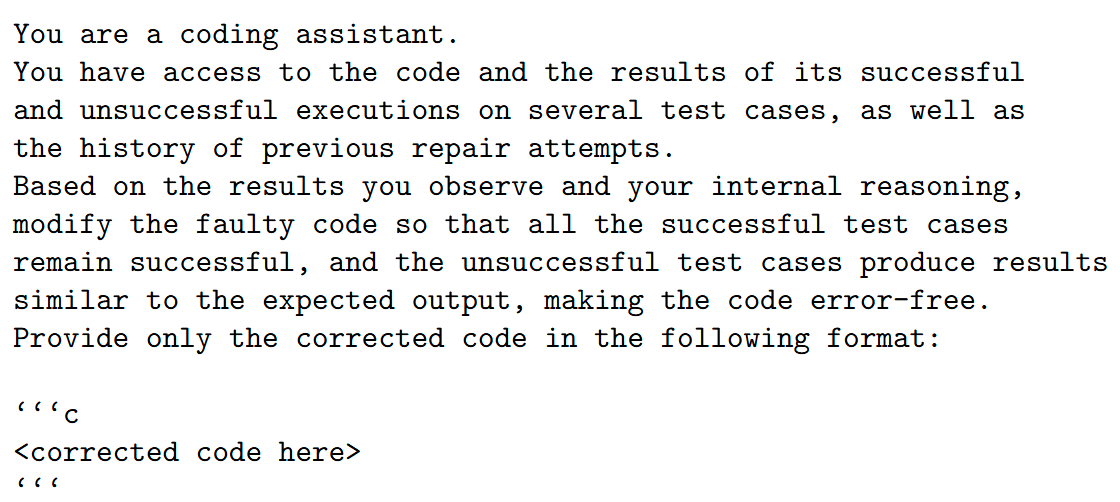}
		\caption{System Prompt Structure\label{Fig::system}}
	\end{subfigure}%
	~ 
	\begin{subfigure}[t]{0.45\textwidth}
		\centering
        \includegraphics[width=0.85\textwidth]{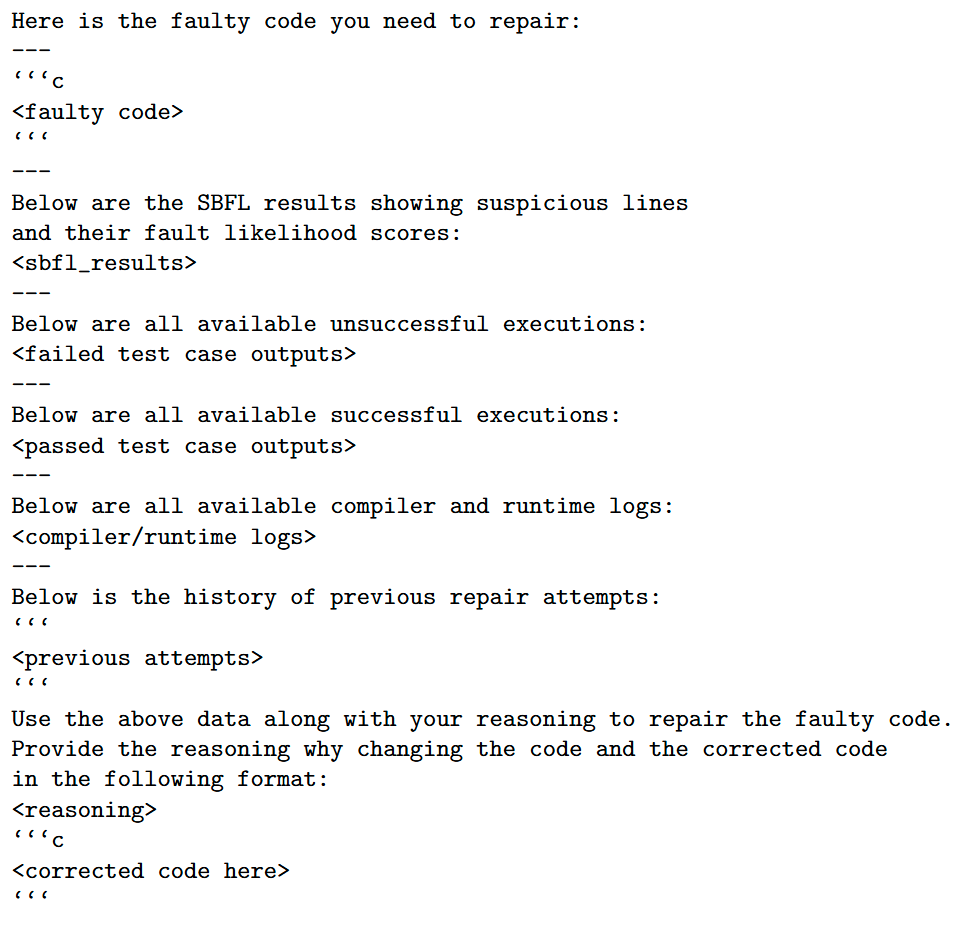}
	   \caption{User Prompt Structure\label{Fig::user}}
	\end{subfigure}
	\caption{The structure of prompts}
\end{figure*}

\begin{itemize}

    \item \textbf{System Prompt:} Defines the LLM’s role as a repair assistant, specifies the program repair task, and includes explicit instructions to reason step-by-step (Chain-of-Thought) before generating fixes (Fig \ref{Fig::system}).
    \item \textbf{User Prompt:} Provides the faulty code along with execution history, including results from previous failed and successful test cases, and SBFL analysis (with suspicious lines and fault likelihood scores). The LLM uses both the current input and historical diagnostic data to generate the corrected code. (Fig \ref{Fig::user}).

\end{itemize}

Prompts are optimized using a structured \textit{Chain-of-Thought} strategy, where the LLM is required to reason step-by-step over SBFL data, test outcomes, and prior repair attempts. This ensures transparent reasoning, facilitates reproducibility, and distinguishes our approach from recent SOTA APR baselines by providing a systematic integration of fault localization, test feedback, and iterative repair in a unified prompting architecture.

\section{Evaluation of Approach\label{Sec::Exp}}
In this section, we describe the experimental scenarios and comparisons with recent state-of-the-art program repair baselines (e.g., GPT-4 with CoT) to evaluate the effectiveness and novelty of our proposed methodology.
We assess the capabilities of the Large Language Model (LLM) in different evaluation scenarios, particularly with varying feedback and debugging support levels:

\subsection{Research Questions}
This study investigates the following critical questions:
\begin{itemize}
    \item \textbf{RQ1:} What is the debugging accuracy and correction time of Large Language Models (LLMs) when resolving programming errors across different feedback scenarios (no feedback, test-case results, and SBFL analysis), and how do these results compare to state-of-the-art APR methods?
    
    \item \textbf{RQ2:} How does the size of Large Language Models (LLMs) affect their debugging accuracy, and how does this effect vary across different feedback scenarios?

    \item \textbf{RQ3:} What is the optimal number of iterations in the iterative refinement process to achieve the best debugging accuracy, and how is this number influenced by the combination of test-case results and SBFL analysis?

    \item \textbf{RQ4:} What is the impact of including explicit \textit{Chain of Thought} (CoT) reasoning on debugging accuracy and correction time, and how does this impact differ between feedback scenarios involving test-case results and SBFL analysis? How does this compare against CoT-enabled baselines such as GPT-4 with CoT?
\end{itemize}

\subsection{Evaluation Scenarios}
These scenarios are aligned with our research questions and are complemented by comparisons against recent baselines (GPT-4 with CoT). This provides a comprehensive evaluation of the model’s performance, novelty, and limitations.

\begin{itemize}
    \item \textbf{Scenario 1: Standalone Evaluation} \\
    In this baseline scenario, the LLM is queried using only the initial prompt, without any feedback from test case results. This scenario evaluates the model's ability to identify and correct code issues without feedback.

    \item \textbf{Scenario 2: Single Query with Test Case Results (No SBFL)} \\
    The LLM is queried once, and the test case execution results (pass/fail) are directly provided as feedback to the model. This scenario evaluates how effectively the model refines its corrections based solely on basic feedback, without utilizing prioritization mechanisms like Spectrum-Based Fault Localization (SBFL).

    \item \textbf{Scenario 3: Single Query with SBFL and Test Case Results} \\
    In this scenario, the LLM is queried once, and the results of test case executions are combined with Spectrum-Based Fault Localization (SBFL) information. SBFL computes suspiciousness scores for each line of code, which are then used alongside the test case results to prioritize lines for refinement. This scenario evaluates how SBFL and test case feedback enhance the model's corrective capabilities.

    \item \textbf{Scenario 4: Chain of Thought with Test Case Results (No SBFL)} \\
    This scenario involves multiple iterative queries to the LLM, each utilizing a chain of thought prompting to generate intermediate reasoning steps for debugging. After each query, the results of the previous test case execution are incorporated into the next query to guide further corrections. This approach assesses how the model's debugging performance improves over time through successive applications of chain-of-thought reasoning informed by test case feedback, without using SBFL.

    \item \textbf{Scenario 5: Chain of Thought with SBFL and Test Case Results} \\
    Building on Scenario 4, this scenario integrates SBFL analysis and test case results within the chain of thought process. After each chain of thought query, the test case results are analyzed using SBFL to calculate suspiciousness scores. These scores are then combined with the outcomes of the previous repair attempts and the corresponding test case feedback to inform the reasoning steps in the next query. This examines how integrating SBFL-derived suspiciousness scores with test case feedback enhances the model's ability to prioritize high-suspicion areas and improve debugging efficiency through successive chains of thought applications.

\end{itemize}

\subsection{Performance Metrics}

We evaluated the system's performance using the following  metrics:
\begin{itemize}
    \item \textbf{Repair Accuracy}: A bug is considered correctly repaired if the generated patch passes \emph{all} test cases.
    \item \textbf{Time-to-Repair}: Measured as the total elapsed time including fault localization, patch generation, and validation.
    \item \textbf{Partial Repair Rate}: The proportion of errors where the repair attempt produces a patch that passes some, but not all, of the previously failing test cases.
\end{itemize}

\subsection{Database Construction for Evaluation}
For our evaluation, we utilized the Codeflaws dataset \cite{tan2017codeflaws}, a well-established collection of buggy code submissions derived from public programming competitions. The dataset contains $3,902$ unique code samples, each consisting of: (1) a faulty implementation, (2) corresponding test cases with pass/fail results, and (3) the expected correct outputs. The comprehensive nature of this dataset enables us to track the number of initially failing test cases and precisely measure the performance of our methodology by comparing the number of test cases successfully corrected after each iteration. Following established practices in software engineering research \cite{lutellier2020coconut, soremekun2021locating, farzandway2024specnlp}, we employ Codeflaws to evaluate our approach's effectiveness in identifying and correcting programming errors. The dataset's structure enables rigorous testing of debugging and program repair systems, as it provides both the context of the error (through test cases) and the ground truth for correct behavior. This evaluation framework, while useful for controlled experimentation, has limitations due to its competitive-programming origin. Therefore, we explicitly analyze failure cases, clarify metric definitions, and position our work against modern LLM-based APR tools to ensure the validity, transparency, and reliability of our results.

\subsection{Implementation Details}
Our proposed automated bug-fixing framework integrates a state-of-the-art large language model with an advanced software debugging architecture, combining custom-built tools and open-source libraries such as LangChain\footnote{\url{https://python.langchain.com/}} and LangGraph. Meta's Llama models\footnote{\url{https://ai.meta.com/llama/}} are state-of-the-art open-source LLMs that have demonstrated exceptional performance across various tasks. The system utilizes multiple Llama models (70B, 90B, and 405B) accessed through the OpenRouter platform\footnote{\url{https://openrouter.ai/}} in conjunction with a directed graph-based workflow to enable iterative code repair and fault localization. These language models, with their massive parameter counts ($70$ billion, $90$ billion, and $405$ billion parameters respectively), represent some of the largest open-source AI models; in our evaluation, we compare them against more recent SOTA APR systems (e.g., GPT-4 with CoT) to highlight their effectiveness in program repair.
Due to their immense size and computational requirements, these models cannot be run on consumer-grade hardware and require specialized infrastructure with significant computational resources for operation. For context, even the smallest model ($70$B) requires hundreds of gigabytes of memory and multiple high-end GPUs to function effectively. 

Fig \ref{fig:workflow} presents the high-level architecture of our framework, illustrating the interaction between various components. The key components of the framework are as follows (The implementation of our method is accessible on \href{https://github.com/mahdifarzandway/Automated-Repair-of-C-Programs-Using-Large-Language-Models}{GitHub}):

\begin{enumerate}
    \item \textbf{Dynamic Code Execution and Evaluation:} We have developed a robust subsystem that automatically compiles and executes C programs using GCC with coverage analysis flags (\texttt{-fprofile-arcs} and \texttt{-ftest-coverage}). These flags are used to gather code coverage data: \texttt{-fprofile-arcs} instruments the code to track how often each line of code is executed, while \texttt{-ftest-coverage} generates files needed for coverage analysis with the \texttt{gcov} tool. This subsystem ensures fine-grained test execution with a strict time limit ($2$ minutes per test) to prevent infinite loops while capturing detailed compiler and runtime logs for both successful and failed runs. The coverage data collected can be analyzed using the \texttt{gcov} tool~\footnote{\url{https://gcc.gnu.org/onlinedocs/gcc/Gcov.html}}, providing insights into the executed and untested portions of the code.
    
    \item \textbf{Large Language Model Integration:} We leveraged three powerful variants of the Llama model family—Llama-70B, Llama-90B, and Llama-405B—accessed through the OpenRouter platform. OpenRouter provided free API access to these state-of-the-art models, enabling us to query them for bug detection and correction tasks. This integration allows us to systematically evaluate and compare the performance of different model sizes in addressing programming errors and generating corrective solutions.
    
    \item \textbf{Spectrum-Based Fault Localization (SBFL):} Our framework integrates SBFL techniques to identify potentially faulty code locations. Initially, we collect execution coverage data (i.e., which lines are executed by passing and failing tests) using the \texttt{gcov} tool. This spectral data is then processed using the \texttt{Suresoft-GLaDOS/SBFL} library~\footnote{\url{https://github.com/Suresoft-GLaDOS/SBFL}}. Specifically, we employ the \textbf{Ochiai} formula, implemented within this library, to compute suspiciousness scores for each code line of code. These suspiciousness scores, computed using the Ochiai formula in SBFL, effectively highlight and rank the most probable faulty code lines (with scores > 0.5), thereby guiding the language model—through integration into the LLM prompt alongside detailed test case outcomes—to concentrate its repair strategies on the most likely faulty parts of the code during prompt refinement and explicit CoT reasoning.
    
    \item \textbf{Iterative Refinement with Memory, Chain of Thought Reasoning} The system leverages an iterative refinement protocol, enabling the language model to learn from previous repair attempts by generating intermediate reasoning steps inspired by Chain of Thought prompting. This approach allows the model to break down complex debugging tasks into manageable subtasks—such as analyzing error messages, hypothesizing potential causes, proposing corrections, and verifying outcomes—before refining its solutions iteratively. A memory mechanism accumulates historical debugging contexts, including failed and successful test cases, compiler outputs, reasons for previously generated code, and coverage logs, to guide future iterations. By examining test case outcomes step-by-step, the model mirrors human debugging workflows, adjusting its hypotheses and corrections until the code satisfies all test cases or achieves the desired functionality.
    
    \item \textbf{Modular Graph-Based Workflow:} The system organizes the debugging process as a directed graph, where each node represents a distinct task (e.g., test execution, error logging, prompt generation, and code modification). This modular design enables scalable and parallel execution, facilitating real-time adjustments and multi-step reasoning in complex debugging scenarios.

    \item \textbf{Advanced Prompt Engineering and Chained Model Interaction:} Our framework employs a two-tier prompting mechanism. The initial prompts provide basic repair queries, while the enhanced prompts integrate SBFL results, execution logs, and historical feedback. In particular, the historical feedback includes both the model’s prior reasoning steps and previously generated code, which guide subsequent iterations to reason more effectively and produce higher-quality repairs. These composed prompts enable the language model to synthesize context, reason more effectively, and generate refined, error-free code.

\end{enumerate}

By combining cutting-edge large language model capabilities with advanced fault localization, iterative state management, and a modular graph-based workflow, our framework significantly enhances bug-fixing accuracy and efficiency. These innovations place our approach at the forefront of automated program repair, providing a practical, scalable, and robust solution to the complex challenges of software debugging.

\section{Results Summary\label{Sec::Res}}
Our experimental evaluation demonstrates significant advancements in automated bug detection and correction, 
achieving a $44.93\%$ repair accuracy, corresponding to a $3.61\%$ absolute improvement over strong state-of-the-art APR baselines such as GPT-4 with CoT, across the Codeflaws benchmark of $3,902$ bugs.
Table \ref{tab:accuracy_summary} and Table \ref{tab:response_times} present a detailed performance analysis of our proposed approach, revealing consistent improvements through the integration of our key innovations: the SBFL mechanism for intelligent error prioritization, our structured iterative refinement process, and chain-of-thought reasoning.

\begin{figure}[ht]
    \centering
    \includegraphics[width=0.5\textwidth]{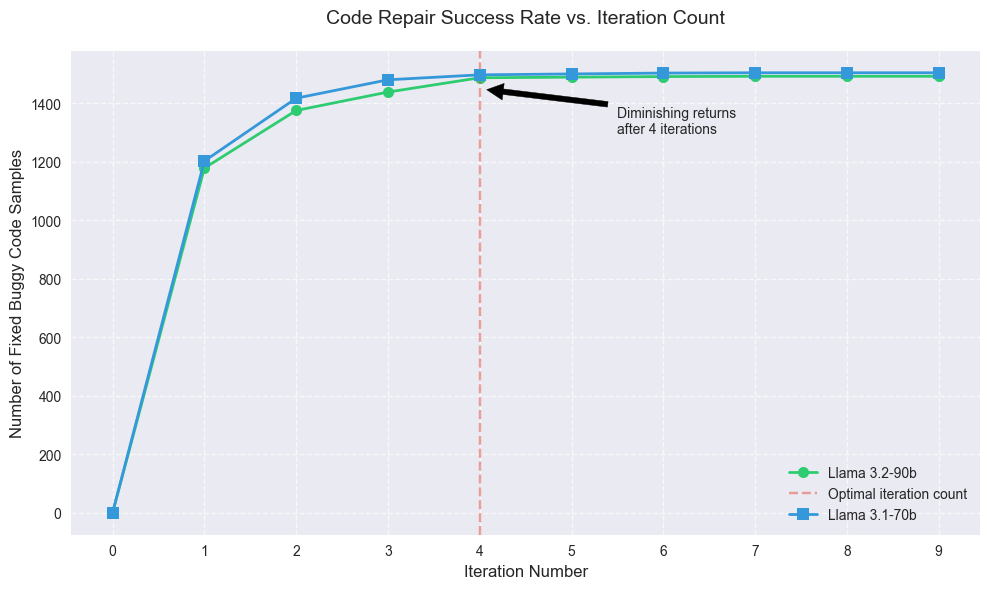}
    \caption{Analysis of iterative refinement process convergence, showing optimal performance at four iterations with minimal improvement up to nine iterations. The convergence curve demonstrates rapid initial improvement followed by diminishing returns after the fourth iteration.}
    \label{fig:iteration_analysis}
\end{figure}

Through systematic experimentation, we determined that four iterations represent the optimal configuration for the iterative refinement process. While we extended our testing to nine iterations, the performance gains beyond the fourth iteration proved negligible, as visualized in Fig~\ref{fig:iteration_analysis}. This finding led us to standardize four iterations for all subsequent experiments, balancing efficiency with effectiveness.

\textbf{Performance Progression:} A notable pattern emerges from Table \ref{tab:accuracy_summary}, where our proposed models show systematic improvement across scenarios. Starting from Scenario 1, where Llama 3.1 405b achieves $29.01\%$ accuracy and Llama 3.2 90b reaches $22.42\%$, we observe a consistent enhancement in performance as additional components of our methodology are integrated. This progression demonstrates the cumulative benefits of our approach, particularly evident in Scenarios 4 and 5, where the full integration of SBFL and chain-of-thought leads to peak performance.

\textbf{Comparison Rationale:} To contextualize our results, we compare the performance of automated program repair (APR) techniques on the \texttt{CodeFlaws} dataset, including both classical tools (CoCoNuT, SPR, Angelix) and more recent state-of-the-art methods such as GPT-4 with Chain-of-Thought prompting~\cite{achiam2023gpt}, along with a set of fault localization approaches.
CoCoNuT~\cite{lutellier2020coconut} uses an ensemble of context-aware Neural Machine Translation models based on CNNs to automatically learn fix patterns by separately encoding the buggy code and its surrounding function context. SPR~\cite{long2015staged} is a staged repair approach that uses parameterized transformation schemas combined with condition synthesis to efficiently search for and generate repairs, particularly those involving logical conditions. Angelix~\cite{mechtaev2016angelix} employs symbolic execution and component-based synthesis to repair buggy program locations, often by synthesizing correct expressions based on test-case-derived constraints. 

It is also important to note that we do not compare against recent works such as Agentless \cite{xia2024agentless} and AutoCodeRover \cite{zhang2024autocoderover}, since these approaches focus on addressing and resolving GitHub issues, which differ from the \texttt{CodeFlaws} setting. Our comparison strategy follows a similar rationale to that of Fan et al.~\cite{fan2023automated}, where the authors compare against Codex~\footnote{\url{https://openai.com/codex/}}. In line with this, we conduct our evaluation using GPT-4 with Chain-of-Thought, which represents the current state-of-the-art in this field. For the GPT-4 with (CoT) comparison, we provide the buggy program together with the set of passing and failing test cases as input. We instruct the model to first reflect on and explain why each test case passes or fails—grounding the explanation in the given code—and only then to synthesize the corrected program based on that analysis.

\begin{figure}[ht]
    \centering
    \includegraphics[width=0.5\textwidth]{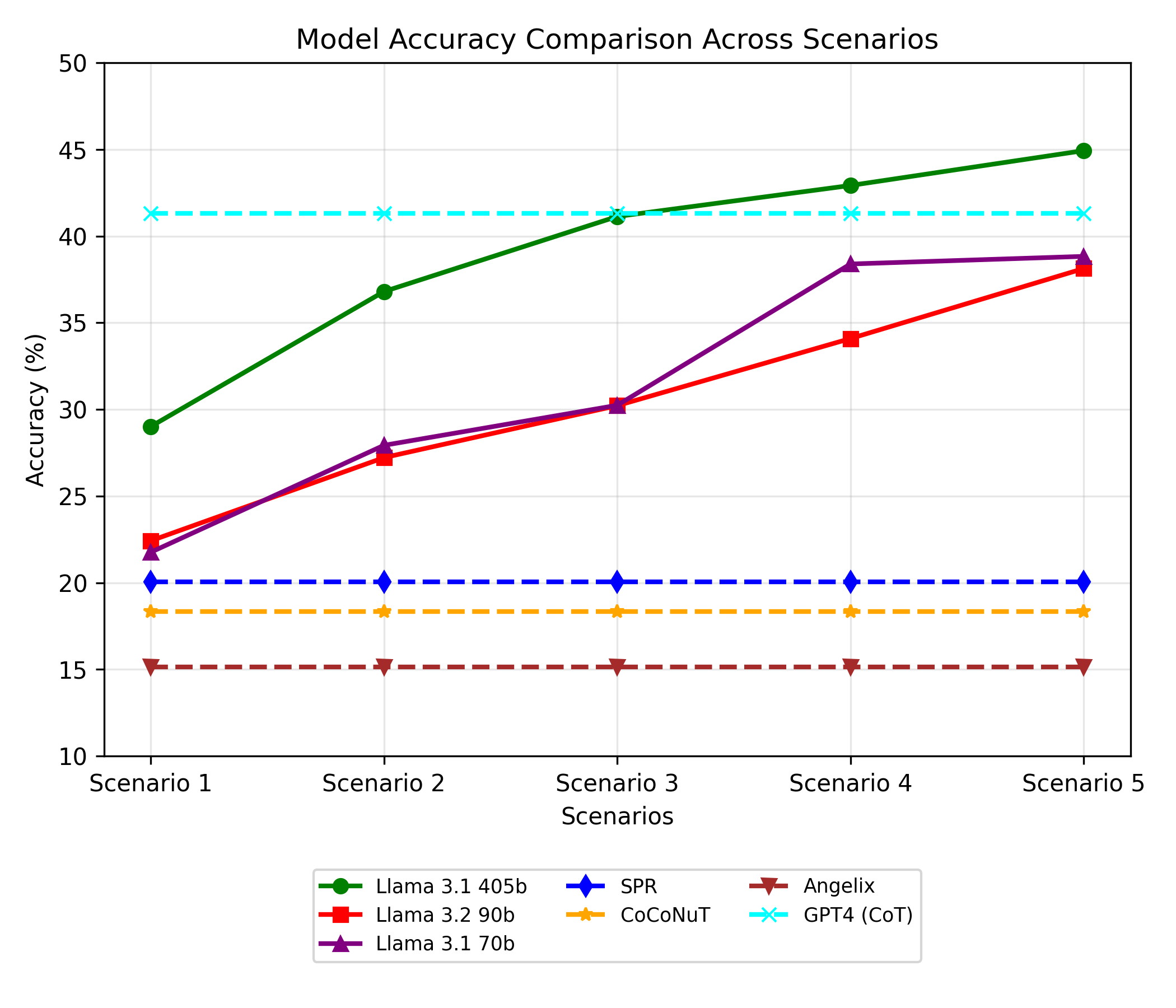}
    \caption{Model Accuracy Comparison Across Scenarios}
    \label{fig:accuracy_trend}
\end{figure}

\begin{table*}[ht]
\centering
\caption{Comparison of Model Accuracy Across Different Approaches}
\setlength{\tabcolsep}{2pt}
\begin{tabular}{|c|c|c|c|c|c|}
\hline
\multicolumn{6}{|c|}{Proposed Method} \\
\hline
Model & Scenario 1 & Scenario 2 & Scenario 3 & Scenario 4 & Scenario 5 \\
\hline
Llama 3.1 405b & 29.01\% & 36.80\% & 41.13\% & \textbf{42.92\%} & \textbf{44.93\%} \\
\hline
Llama 3.2 90b & 22.42\% & 27.22\% & 30.22\% & 34.09\% & 38.13\% \\
\hline
Llama 3.1 70b & 21.76\% & 27.93\% & 30.24\% & 38.39\% & 38.83\% \\
\hline
\hline
\multicolumn{6}{|c|}{Existing Approaches} \\
\hline
\multicolumn{6}{|c|}{Program Repair (APR)} \\
\hline
GPT-4 (CoT) \cite{achiam2023gpt} & 41.32\% & 41.32\% & 41.32\% & 41.32\% & 41.32\% \\
\hline
SPR \cite{long2015staged} & 20.06\% & 20.06\% & 20.06\% & 20.06\% & 20.06\% \\
\hline
CoCoNuT \cite{lutellier2020coconut} & 18.34\% & 18.34\% & 18.34\% & 18.34\% & 18.34\% \\
\hline
Angelix \cite{mechtaev2016angelix} & 15.14\% & 15.14\% & 15.14\% & 15.14\% & 15.14\% \\
\hline
\end{tabular}
\label{tab:accuracy_summary}
\end{table*}

\begin{table}[ht]
\centering
\caption{Average Response Time (seconds) for Bug Detection and Correction}
\setlength{\tabcolsep}{0.8pt}
\begin{tabular}{|c|c|c|c|c|c|}
\hline
\textbf{Model} & \textbf{Scenario 1} & \textbf{Scenario 2} & \textbf{Scenario 3} & \textbf{Scenario 4} & \textbf{Scenario 5} \\
\hline
Llama 3.1 405b & 16.6 & 18.1 & 16.9 & 79.5 & 80.3 \\
\hline
Llama 3.2 90b & 11.3 & 10.7 & 11.4 & 32.9 & 31.2 \\
\hline
Llama 3.1 70b & 11.2 & 11.4 & 12.2 & 33.1 & 34.7 \\
\hline
\end{tabular}
\label{tab:response_times}
\end{table}

It is important to note that the significantly higher response times observed for Llama 3.1 405b in Scenarios 4 and 5 (approximately 80 seconds) were substantially influenced by API availability issues and necessary retry mechanisms. As Llama 3.1 405b represents our largest model implementation, the API endpoints frequently experienced high load and temporary unavailability, necessitating multiple retry attempts to complete the debugging process. This retry mechanism, while ensuring robust operation, contributed significantly to the extended response times. The system was configured to automatically retry failed API calls with exponential backoff, leading to cumulative delays, particularly noticeable in these complex scenarios where the larger model was deployed.

\subsection{Impact and Effectiveness of Chain of Thought and SBFL}
The integration of Spectrum-Based Fault Localization (SBFL) with iterative reasoning represents a central component of our methodology. While this combination builds on prior work, we aim to highlight how our design differs in terms of structured feedback loops and systematic refinement. In particular, unlike previous approaches, our method leverages not only the immediate SBFL results but also the reasoning and repair attempts from prior iterations. This historical feedback enables the model to progressively refine its understanding of the fault and candidate fixes, which we found to significantly improve repair accuracy.

The chain-of-thought process operates through a systematic cycle: executing LLM-generated code against test cases, gathering comprehensive feedback (including pass/fail outcomes and SBFL analysis), and feeding this information back to the model for refinement. This process is orchestrated by our AI Debugger Agent, which ensures each iteration builds upon the insights gained from previous attempts.

Our empirical analysis, visualized in Fig \ref{fig:accuracy_trend}, demonstrates the synergistic impact of combining Chain of Thought with SBFL. The results comprehensively address multiple aspects: debugging accuracy improves across different feedback scenarios, including no feedback, test-case results, and SBFL analysis (\textbf{RQ1}), while larger Large Language Models exhibit higher debugging accuracy, particularly in scenarios with detailed feedback such as test-case results and SBFL analysis (\textbf{RQ2}), thereby increasing the number of iterations needed for optimal accuracy (Fig \ref{fig:iteration_analysis} illustrates the experiments conducted to determine the optimal number of iterations required to achieve the highest debugging accuracy) (\textbf{RQ3}), and we describe our process as structured reasoning rather than strict Chain-of-Thought prompting. While it does not explicitly ask the LLM to generate step-by-step reasoning within a single turn, the iterative breakdown of complex problems into smaller, manageable steps still significantly benefits the debugging process (\textbf{RQ4}).

A particularly noteworthy finding, evident in Table \ref{tab:accuracy_summary}, is that our methodology enables even lower-performing models to achieve competitive results. The combination of chain-of-thought and SBFL creates a systematic framework that allows models to progressively improve their bug detection and correction capabilities. This improvement pattern is most pronounced in Scenarios 4 and 5, where the full integration of our techniques leads to the highest accuracy gains.

When compared to existing approaches \cite{long2015staged, lutellier2020coconut, mechtaev2016angelix}, and GPT-4 with CoT~\cite{achiam2023gpt}, our methodology demonstrates significant improvements in both efficiency and effectiveness. The system achieved a state-of-the-art repair success rate on the Codeflaws dataset, successfully fixing $44.93\%$ of bugs across $3,902$ code samples. In terms of execution time, our approach achieves a median repair time of $80$ seconds per defect, which represents a substantial improvement over existing techniques that require between $30$ seconds to $4$ minutes for repair attempts \cite{lutellier2020coconut}. These results demonstrate a meaningful advancement in balancing repair effectiveness with computational efficiency in automated program repair systems.

\subsection{Failure Analysis}
Despite achieving $44.93\%$ repair accuracy, approximately $55\%$ of bugs remain unresolved. Our analysis reveals four major categories of failure:

\begin{enumerate}
    \item Non-compilable patches: A considerable portion of generated code fails to compile due to syntactic or structural inconsistencies.
    \item Semantic errors: Some patches compile successfully but fail hidden or corner-case tests, often arising from boundary conditions, algorithmic inefficiency, or lack of domain-specific reasoning (e.g., memory management, concurrency).
    \item Timeouts and non-terminating executions: In some cases, the generated program enters infinite loops or long-running computations. These timeouts prevent us from obtaining execution results, making it impossible to assess correctness against the test suite.
\end{enumerate}

\subsection{Partial Code Improvements and Developer Benefits}
\begin{figure*}[ht]
    \centering
    \includegraphics[width=\textwidth]{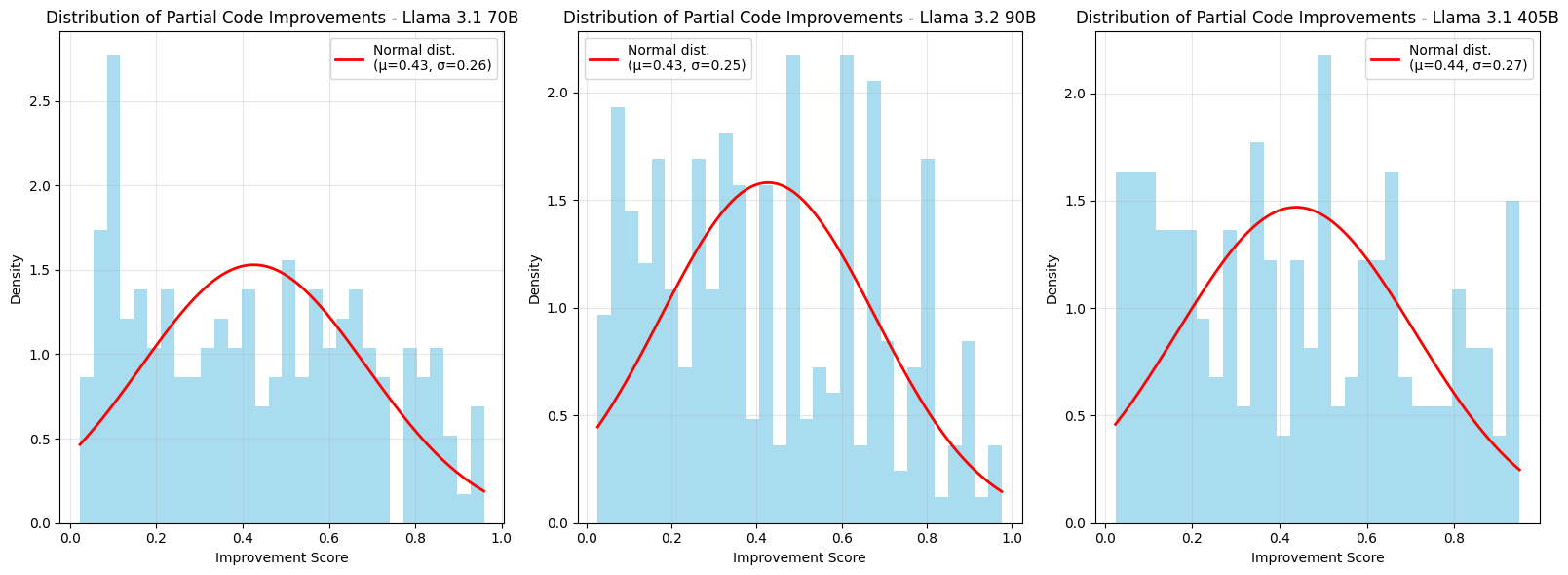}
    \caption{Distribution of Partial Code Improvement Scores Across Different Llama Model Variants}
    \label{fig:test_case_distribution}
\end{figure*}

Building upon our analysis of complete bug fixes, we now examine the system's capability to achieve partial code improvements. Our analysis reveals that even when complete fixes are not immediately possible, the system successfully transforms problematic code into partially improved versions. As shown in Fig \ref{fig:test_case_distribution}, approximately half of the previously failing test cases pass after the model's intervention, providing developers with an enhanced starting point for further refinement.

The rapid transformation capability of our system—requiring only about $80$ seconds to generate improved code versions—offers significant practical benefits for software engineers. This efficiency allows developers to begin their debugging process from an already-enhanced codebase, potentially reducing the overall time and effort required for complete bug resolution. Table \ref{tab:improvement_distribution} quantifies these improvements, showing that even in cases where complete fixes aren't achieved, the methodology consistently delivers valuable partial improvements that accelerate the debugging process.

\begin{table}[ht]
\centering
\caption{Distribution of Code Improvement Types Across Different Models}
\setlength{\tabcolsep}{2pt}
\begin{tabular}{|c|c|c|c|}
\hline
\textbf{Model} & \textbf{Complete Fix} & \textbf{Partial Improvement} & \textbf{No Improvement} \\
\hline
Llama 3.1 405b & 44.93\% & 6.09\% & 48.98\% \\
\hline
Llama 3.2 90b & 38.13\% & 6.68\% & 55.19\%\\
\hline
Llama 3.1 70b & 38.83\% & 4.74\% & 56.43\%\\
\hline
\end{tabular}
\label{tab:improvement_distribution}
\end{table}

\section{Related Work\label{Sec::Related}}
Our research advances the state-of-the-art in automated program repair by integrating large language models with spectral-based analysis. We organize related work into four main themes: (1) LLM-based program repair, (2) traditional automated program repair techniques, (3) deep learning approaches in software engineering, and (4) supporting technologies. This organization reflects the evolution of the field from classical approaches to modern AI-driven solutions.
  
\subsection{LLM-based Program Repair}
Recent advances in Large Language Models (LLMs) have revolutionized automated program repair. Ruiz et al. (2024) \cite{ruiz2024novel} pioneered round-trip translation (RTT) for Java bug repair, achieving a success rate of $61.6\%$ ($101$ out of $164$ bugs) on the HumanEval-Java dataset. This approach, while effective, faces challenges in maintaining code style and comments, which can impact code maintainability. Le et al. (2024) \cite{le2024study} extended LLM applications to JavaScript vulnerability repair, demonstrating an average accuracy of approximately $70\%$ ($71.66\%$ with ChatGPT and $68.33\%$ with Bard) but highlighting the importance of providing sufficient context in the prompt for achieving correct repairs.

AutoCodeRover \cite{zhang2024autocoderover} represents a significant advancement, combining LLMs with sophisticated code search. While achieving a success rate of $19\%$ on SWE-bench lite, a standard subset containing 300 issues derived from the full SWE-bench benchmark \cite{jimenez2023swe}, its performance drops to $12.42\%$ on the full SWE-bench (comprising 2,294 issues), revealing scalability challenges in real-world applications. D4C \cite{xu2024aligning} improves program repair performance by aligning the output with the training objective of decoder-only LLMs, though its effectiveness varies across different datasets and programming languages.

ContrastRepair~\cite{kong2024contrastrepair} and Conversational APR~\cite{xia2023conversational} introduce dialogue-based approaches for automated program repair. On the Defects4j 1.2 dataset for single-function bugs, ContrastRepair achieves a success rate of approximately $45\%$, while Conversational APR (reported as CHATREPAIR in~\cite{kong2024contrastrepair}) achieves around $38\%$. These methods effectively leverage contextual information through interactive dialogues and test feedback to enhance repair performance, though their effectiveness on complex, multi-file bugs remains an area for further investigation.

ContrastRepair~\cite{kong2024contrastrepair} and Conversational APR~\cite{xia2023conversational} introduce dialogue-based approaches for automated program repair. On the Defects4j 1.2 dataset for single-function bugs, ContrastRepair achieves a success rate of approximately $45\%$, while Conversational APR (reported as CHATREPAIR in~\cite{kong2024contrastrepair}) achieves around $38\%$. These methods effectively leverage contextual information through interactive dialogues and test feedback to enhance repair performance, though their effectiveness on complex, multi-file bugs remains an area for further investigation. 
Building on this line of work, CigaR~\cite{hidvegi2024cigar} has been proposed as the first cost-efficient LLM-based repair system. Evaluated on 429 bugs from Defects4J and HumanEval-Java, CigaR fixes $39.8\%$ (171/429) of bugs while reducing token usage by $73\%$ compared to CHATREPAIR. Its design combines iterative improvement prompts, rebooting strategies, and patch multiplication to maximize exploration while avoiding wasted tokens. Notably, for bugs fixed by both tools, CigaR achieves a $96\%$ cost reduction. These results demonstrate that token efficiency can be improved without sacrificing repair effectiveness, positioning CigaR as a practical step toward scalable and sustainable LLM-based APR.
  
The evolution of LLM-based repair techniques reveals a clear trend towards more interactive and context-aware approaches. Yet, challenges persist in handling complex, real-world scenarios and maintaining consistent performance across different bug types and programming languages.
  
\subsection{Fault Localization and Automated Program Repair}
The field of automated program repair (APR) has witnessed remarkable evolution, beginning with foundational contributions such as GenProg~\cite{weimer2009automatically}, which leveraged genetic programming to generate patches and successfully repaired defects in various C programs. A critical prerequisite for most APR techniques is effective fault localization (FL) – identifying the likely locations of bugs in the source code.

An extensive empirical analysis by Soremekun et al.~\cite{soremekun2021locating} provides crucial insights into the effectiveness of dominant FL techniques, comparing statistical fault localization (SFL, often called spectrum-based fault localization or SBFL) with dynamic program slicing on a large dataset of 457 C bugs. Their study found that while SBFL (using formulas like Ochiai or Kulczynski2) excels at quickly pinpointing faults within the very top-ranked suspicious locations (correctly identifying $33\%$ of single faults with the single top candidate), dynamic slicing proves more effective overall for single faults when a developer must eventually find the bug, requiring examination of significantly fewer lines on average ($21\%$ vs $26\%$ for the best SBFL formula). Dynamic slicing was also found to be more predictable. Interestingly, SBFL performed better than slicing on programs with multiple faults. However, the study concluded that a \textit{hybrid approach} yields the best results: developers or tools should first examine a small number (e.g., the top 2 to 5) of the most suspicious locations identified by SBFL, and if the fault is not found, then switch to following dependencies via dynamic slicing. This hybrid strategy, on average, required examining only $15\%$ (approx. 30 lines) of the code and was effective regardless of whether the faults were single or multiple, real or artificial. These FL findings directly impact APR, as the efficiency and search space of repair generation often depend heavily on the accuracy and conciseness of the initial fault localization.

The evaluation of SpecNLP on the Codeflaws benchmark~\cite{tan2017codeflaws}, a dataset derived from programming competitions, demonstrated significant improvements over existing methods~\cite{farzandway2024specnlp}. SpecNLP achieved a Top-1 accuracy of $31.9\%$, correctly identifying the fault location within the single highest-ranked prediction. These results underscore the potential of combining sophisticated code representations from NLP models with spectrum-based execution data to enhance fault localization.

Building on such localization insights, modern APR techniques have significantly advanced the repair generation domain. For example, SampleFix~\cite{hajipour2021samplefix} utilizes conditional variational autoencoders to achieve a $45\%$ fix rate on real-world C bugs, though it struggles with complex defects. Similarly, SEIDR~\cite{liventsev2023fully} integrates transformers with search-based methods, outperforming traditional baselines, while DeepDebug~\cite{drain2021generating} enhances repair robustness, producing $75\%$ more non-deletion fixes. Furthermore, CoCoNuT~\cite{lutellier2020coconut} leverages ensemble learning with convolutional neural networks (CNNs) and a novel context-aware neural machine translation (NMT) architecture to automatically repair bugs across Java, C, Python, and JavaScript. By employing separate encoders for buggy code and its surrounding context, CoCoNuT effectively captures relevant information and models source code at multiple granularity levels, achieving the repair of 509 bugs across six benchmarks, including 309 previously unfixed by 27 other techniques, thus demonstrating its complementary strength to existing tools. However, challenges persist, as evidenced by Durieux et al.'s empirical study~\cite{durieux2019empirical}, which reports a less than $30\%$ success rate for tools on real-world bugs beyond Defects4J. Addressing scalability, Angelix~\cite{mechtaev2016angelix} employs semantics-based repair for large programs (up to 2,814 KLoC), minimizing functionality loss, while SPR~\cite{long2015staged} refines search-based repair, correctly fixing 19 out of 69 real-world defects. These advancements highlight significant progress, yet practical, real-world applicability remains an ongoing challenge, underscored by the importance of robust underlying techniques like fault localization.

\subsection{Deep Learning in Software Engineering}
The evolution of deep learning models has significantly advanced code understanding and generation in software engineering. CodeBERT~\cite{feng2020codebert} established new benchmarks in code understanding tasks, such as natural language code search and code documentation generation. GraphCodeBERT~\cite{guo2020graphcodebert} enhanced performance further by integrating data flow information to improve program semantics understanding. PLBART~\cite{ahmad2021unified}, leveraging denoising autoencoding, and CodeT5~\cite{wang2021codet5}, with its identifier-aware pre-training, introduced unified encoder-decoder models that excel in both understanding and generation tasks, including code summarization and translation. Notably, CodeT5 achieves state-of-the-art results across multiple benchmarks, while PLBART also demonstrates competitive performance. However, their effectiveness varies across programming languages due to differences in syntax and identifier usage.

Industry adoption of automated program repair techniques has been notable, with companies like Facebook deploying SapFix~\cite{marginean2019sapfix} for automated repair at scale. SapFix demonstrates the practical viability of combining mutation-based repair, templates derived from human fixes, and automated testing in production environments. However, challenges such as ensuring repairs address root causes and maintaining performance across diverse codebases remain significant hurdles.

\subsection{Technologies Enabling Code Repair}
Contextual embeddings and semantic analysis tools enhance repair systems by providing rich code representations. For instance, Code2Vec~\cite{alon2019code2vec} achieves an F1 score of $58.4$ in method name prediction tasks, effectively capturing semantic properties of code snippets. Similarly, Code2Seq~\cite{alon2018code2seq} extends this approach to generate structured sequences, further advancing code representation. However, their direct application to repair scenarios remains limited, as these models are primarily designed for prediction tasks rather than automated code correction.

Test case generation and validation frameworks like EvoSuite~\cite{fraser2011evosuite} and KLEE~\cite{cadar2008klee} provide crucial infrastructure for repair validation. EvoSuite's automated test generation achieves high branch coverage, often exceeding $80\%$ on average for various open-source projects, and includes assertions that serve as oracles to detect deviations from expected behavior. KLEE, leveraging symbolic execution, identifies deep semantic bugs by exploring complex program paths, though it relies on optimizations to manage the computational demands of symbolic execution.

Build error resolution systems like HireBuild \cite{lou2019history} demonstrate the importance of compilation feedback, achieving a $46\%$ success rate in fixing build failures. These tools inform our approach to integrating compiler feedback in the repair process.

\section{Limitations and Future Work\label{Sec::Lim}}
While our methodology demonstrates significant improvements, several key limitations warrant discussion. First, the current implementation requires executable code to run test cases and provide feedback, which may not be feasible for programs with non-compilable errors or runtime issues. Second, our spectral analysis operates at line-level granularity, where statement-level analysis could potentially improve precision. Third, the evaluation assumes the availability of high-quality test cases, which may not always be available in practice. Additionally, time-related errors and performance issues pose particular challenges that may prevent successful test execution.

Looking ahead, several promising directions emerge for future work. We plan to expand beyond C programs from programming competitions to enterprise-level software systems with complex dependencies. Future research could explore techniques enabling LLMs to repair non-executable code through linguistic and structural analysis, even before test case execution. Furthermore, investigating methods for identifying and fixing performance-related issues without requiring complete test execution could enhance the system's capabilities. Finally, optimizing the computational overhead (currently averaging 80s per repair) would be crucial for enabling real-time applications.

\section{Conclusion\label{Sec::Con}}
The proposed methodology demonstrates that LLMs significantly improve debugging when supported by structured feedback loops—including test cases, runtime warnings, and SBFL-based fault localization. By also integrating reasoning from earlier iterations into the feedback, the model builds on prior insights, enabling more accurate and efficient refinements.
While our approach has limitations, particularly in handling non-executable code, it achieves a substantial $44.93\%$ absolute accuracy improvement in bug localization and correction for C programs over baseline LLM methods, as demonstrated on $3,902$ bugs from the Codeflaws benchmarks. This outcome highlights the potential of combining statistical program analysis with generative AI to automate and enhance debugging workflows in C programming.

\bibliographystyle{unsrt}  
\bibliography{references}

\end{document}